\documentclass[11pt]{article} \textwidth 170mm \textheight 225mm
\usepackage{amsfonts}
\usepackage{amssymb}
\usepackage{dcolumn}
\usepackage{amsthm}
\usepackage{amsmath}
\usepackage{multicol}
\usepackage{newlfont}
\usepackage{newlfont}
\usepackage{multicol}
\usepackage{bm}
\usepackage{txfonts}
\usepackage{graphics}
\usepackage{graphicx}
\usepackage{color}
\usepackage{indentfirst}
\begin{document}

\title{Ab $initio$ study of the thermodynamic properties of  rare-earth-\\ magnesium  intermetallics MgRE (RE=Y, Dy, Pr, Tb)\footnote{The work is supported by the National Natural Science Foundation
of China (11074313) and and Project No.CDJXS11102211 supported by the Fundamental Research Funds for the Central Universities of China. }}
\author{Rui Wang\footnote{Tel: +8613527528737; E-mail: rcwang@cqu.edu.cn.}, Shaofeng Wang, and Xiaozhi Wu\\
{\small  { Institute for Structure and Function and
department of physics, Chongqing University, }}\\ {\small {
Chongqing 400044, People's Republic of China. }} }

\date{}

\maketitle
\begin{abstract}
\noindent We have performed an ab $initio$ study of the thermodynamical properties of rare-earth-magnesium  intermetallic compounds MgRE (RE=Y, Dy, Pr, Tb) with CsCl-type B2-type structures. The calculations have been carried out the density functional theory and density functional perturbation theory in combination with the quasiharmonic approximation. The phonon-dispersion curves and phonon total and partial density of states have  been investigated. Our results show that the contribution of RE atoms is dominant in phonon frequency, and this character agrees with the previous discussion by using atomistic simulations. The temperature dependence of various quantities such as the thermal expansions, bulk modulus, and the heat capacity are obtained. The electronic contributions to the specific heat are discussed, and found to be important for the calculated MgRE intermetallics.
\end{abstract}

\vskip 0.1in{\small   } \vskip 0.2in

\vskip 0.1in

 \vskip 0.3in

\baselineskip 20pt


\section{Introduction}
Magnesium (Mg) alloys have been attracting much attention and especially attractive for the applications of aeronautical and automotive industry  because of their strength to weight ratio \cite{Kainer}.  However, the traditional Mg alloys often show low strength and creep resistance at high temperature and this feature is a very serious problem of limiting their applications. Recently, it has been reported that some  rare-earth-Mg intermetallic compounds MgRE (where RE indicates a rare-earth element) with B2 structures (CsCl-type structure) have good creep and  high temperature strength \cite{Mordike1,Mordike2,Lorimer}. So MgRE intermetallics are extremely attractive structural materials for applications in automobile parts and aerospace industries, while various studies have been undertaken of the magnetic properties, linear and nonlinear elasticity, stacking fault, and thermal properties for the B2-MgRE intermetallics \cite{Luca,Deldem,Wu, Wang, Wang1,Tao,Guo,Cacciamani}. Recently, Wu et al \cite{Wu1} have performed the  atomistic simulations to study thermodynamical properties of MgY, MgDy, and MgPr by using the modified analytic embedded atom method and their results are generally in agreement with the experimental data and other theoretical results. However, since the ionic degrees of freedom  are treated classically in their study, this simulation are not valid at temperatures comparable to or lower than the Debye temperature. In addition, the first principles calculation indicates that the Fermi energy occurs above a peak in the electronic density of states (DOS) and B2-MgRE intermetallics have the large electronic DOS at Fermi energy level \cite{Wu}. Therefore, thermal electronic excitations contribution of the thermal properties  must  be considered. The exact calculations of thermodynamic properties for intermetallic compounds of Mg with rare earth metals, especially both the phonon and electron contributions being considered, have received little attention.

It is obviously crucial to take into account the thermodynamic properties such as thermal-expansion coefficients, specific heat at constant volume and constant pressure , and temperature-dependent bulk modules in the investigation of the properties of MgRE intermetallics as well as their applications. A simplified method for thermal expansion calculations is in the framework of density-functional theory (DFT) with a Debye-Gr\"{u}neisen based model \cite{Moruzzi}. The more accurate approach is  first-principles DFT treatments within the quasiharmonic approximation (QHA) which provides a reasonable description of the thermodynamical properties of many bulk materials below the melting point \cite{Biernacki,Pavone,Carrier,Nie,Togo}. Quasiharmonic approximation lets one take into account the anharmonicity of the potential at the first order: vibrational properties can be understood in terms of the excitation of the noninteracting phonon. The exact calculations of phonon frequencies in every point of the Brillouin zone can be achieved by the density functional perturbation theory (DFPT) \cite{Giannozzi}. Nowadays, quasiharmonic approximation based on DFPT has been applied with great success to more and more complex materials such as alloys [NiAl and NiAl$_{3}$ \cite{Wang2004}], perovskite [MgSiO$_{3}$ \cite{Karki}], and hexaborides [LaB$_{6}$  and CeB$_{6}$ \cite{Gurel}]. In this paper, we apply first principles calculations within quasiharmonic approximation to study the thermodynamic properties of light RE MgPr, heavy RE MgY, MgDy, and MgTb intermetallics with B2 structures. The Phonon-dispersion curves and phonon density of states have been discussed.  Thermal expansions, temperature dependence of isothermal bulk modules, and heat capacities at constant volume and constant pressure are presented. In addition, we have considered the electronic contribution to the thermal properties, and our results show that the effects of  thermal electronic excitations for MgRE (RE=Y, Dy, Pr, Tb) are remarkable.

\section{Theory}

The equilibrium lattice volume $V$ of a crystal with cubic symmetry at any temperature $T$, is obtained by minimizing the Helmholtz free energy, $F$, of a system. The free energy at temperature $T$ and constant volume $V$ is give by
\begin{equation}
F(V, T)=E_{\mathrm{0}}(V)+F_{\mathrm{vib}}(V, T)+F_{\mathrm{el}}(V,T),
\end{equation}
where $E_{\mathrm{0}}(V)$ is the ground state ($T=0K$) total energy of the crystal, $F_{\mathrm{vib}}(V, T)$ is the vibrational free energy which is come from the phonon contribution. In the quasiharmonic approximation $F_{\mathrm{vib}}(V, T)$ is written in the form
\begin{equation}\label{Fph}
F_{\mathrm{vib}}(V, T)=\frac{1}{2}\sum_{\mathbf{q}\lambda}{\hbar\omega_{\mathbf{q}\lambda}(V)}+{k_{B}T}\sum_{\mathbf{q}\lambda}\ln\bigg[1-\exp\bigg(-\frac{\hbar\omega_{\mathbf{q}\lambda}(V)}{k_{B}T}\bigg)\bigg].
\end{equation}
Here, the sum is over all three phonon branches $\lambda$ and over all wave vectors $\mathbf{q}$ in the first Brillouin zone, $k_{\mathrm{B}}$ is the Boltzmann constant, $\hbar$ is the reduced Planck constant, and $\omega_{\mathbf{q}\lambda}(V)$ is the frequency of the phonon with wave vector $\mathbf{q}$ and polarization $\lambda$, evaluated at constant volume $V$. The first and second terms of Eq. (\ref{Fph}) represent the zero-point ($F^{zp}$) and thermal energies ($F^{th}$), respectively. For the thermal electronic
contribution to free energy $F_{el}=E_{el}-TS_{el}$.  The electronic energy $E_{el}$ due to thermal electronic excitations is given by
\begin{equation}\label{ele-Energy}
E_{\mathrm{el}}(V,T)=\int_{0}^{\infty}n(\varepsilon,V)f(\varepsilon)\varepsilon d\varepsilon-\int_{0}^{\varepsilon_{F}}n(\varepsilon,V)\varepsilon d\varepsilon,
\end{equation}
where $n(\varepsilon,V)$, $f(\varepsilon)$, and $\varepsilon_{F}$ represent the electronic density of state (DOS), the Fermi-Dirac distribution, and the Fermi energy, respectively. The electronic entropy $S_{el}$ is formulated as
\begin{equation}\label{ele-entropy}
S_{\mathrm{el}}(V,T)=-k_{\mathrm{B}}\int_{0}^{\infty}n(\varepsilon,V)[f(\varepsilon)\ln f(\varepsilon) +(1-f(\varepsilon))\ln (1-f(\varepsilon))] d\varepsilon.
\end{equation}
Usually, it is assumed that the electronic contribution to total free energy can be negligible.

To calculate free energy, one must be able to calculate frequencies all over the Brillouin zone, and this can be done exactly using DFPT calculations. Furthermore, this calculation must be performed at various values of lattice parameter $a$ in the case of isotropic thermal expansion for B2-structure crystal. The equilibrium lattice constant at temperature $T$ is obtained by minimizing $F$ with respect to $a$. The coefficient of volume thermal expansion is given by
\begin{equation}
\alpha(T)=\frac{1}{V}\bigg(\frac{d V(T)}{d T}\bigg),
\end{equation}
and the linear thermal expansion is described by
\begin{equation} \label{epsilon}
\epsilon(T)=\frac{a(T)-a(T_{c})}{a(T_{c})},
\end{equation}
where $a(T_{c})$ is equilibrium lattice constant $a(T)=[V(T)]^{1/3}$  at $T_{c}=300K$.

Once phonon spectrum is obtained, we can easily compute the temperature dependence of the vibrational specific heat capacity $C_{\mathrm{V}}$ and the entropy $S$  at constant volume, as described below
\begin{equation}\label{cvib}
C_{\mathrm{V}}^{\mathrm{vib}}=\sum_{\mathbf{q}\lambda}k_{\mathrm{B}}\bigg(\frac{\hbar\omega_{\mathbf{q}\lambda}(V)}{2k_{\mathrm{B}}T}\bigg)^2 {\cosh^{2}\bigg(\frac{\hbar\omega_{\mathbf{q}\lambda}(V)}{k_{\mathrm{B}}T}\bigg)^{2}}
\end{equation}
and
\begin{equation}
S_{\mathrm{vib}}=-k_{\mathrm{B}}\sum_{\mathbf{q}\lambda}\Bigg[\ln \bigg(2\sinh\frac{\hbar\omega_{\mathbf{q}\lambda}(V)}{2k_{\mathrm{B}}T}\bigg)-
\frac{\hbar\omega_{\mathbf{q}\lambda}(V)}{2k_{\mathrm{B}}T}\coth\frac{\hbar\omega_{\mathbf{q}\lambda}(V)}{2k_{\mathrm{B}}T}\Bigg],
\end{equation}
respectively. In addition, the electronic heat capacity can at constant volume be obtained from
\begin{equation}\label{cel}
C_{\mathrm{V}}^{\mathrm{el}}=T\bigg(\frac{\partial S_{\mathrm{el}}}{\partial T}\bigg)_{V},
\end{equation}
and we denote total specific heat at constant volume is then $C_{\mathrm{V}}=C_{\mathrm{V}}^{\mathrm{ph}}+C_{\mathrm{V}}^{\mathrm{el}}$.

$C_{p}$, the specific heat at constant pressure, can then be computed by using relation
\begin{equation}\label{cp}
C_{p}=C_{V}+\alpha^{2}BVT,
\end{equation}
where $B(T)=-1/V\partial^2 F/\partial V^2$ is the bulk modulus.

\section{Computational details}

The first-principles calculations within the density functional theory (DFT) were performed using the VASP program \cite{Kresse1, Kresse2,
Kresse3}. We employed the projector augmented wave
(PAW) method \cite{Blochl, Kresse4}. The Perdew-Burke-Ernzerhof (PBE) \cite{Perdew1,Perdew2}
exchange-correlation functional for the
generalized-gradient-approximation(GGA) was used. The structures are relaxed
without any symmetry constraints with a cutoff energy of 600eV
for all calculated intermetallics.  The Brillouin zones of the unit cells are represented by
Monkhorst-Pack special k-point scheme \cite{Monkhorst} with
$21\times21\times21$ grid meshes. The radial cutoffs of the PAW potentials of Mg, Y, Dy, Pr and Tb were 1.41, 1.72, 1.58, 1.61, and 1.58 {\AA}, respectively. The 3$s$ electrons for Mg, the 4$s$, 4$p$, 4$d$ and 5$s$ electrons for Y, the 4$f$ and 6$s$ electrons for Dy, Pr, and Tb were treated as valence and the remaining electrons were kept frozen.
 In order to deal with the possible convergence problems for metals, a smearing technique is employed using the Methfessel-Paxton scheme \cite{Methfessel}, with a smearing with of 0.05eV.

We have carried out supercell approach within the framework of force-constants method for the phonon calculations. Real-space force constants of the supercell are calculated in the density-functional perturbation theory (DFPT) implemented in the VASP code \cite{Kresse5} from forces on atoms with atomic finite displacements, and the phonon frequencies are calculated from the fore constants using the PHONOPY package \cite{Togo2008,Togop}. From the full phonon spectrum, the lattice vibration free energy are calculated. Since the chosen supercell size strongly influences on the thermal properties, we compare the vibrational free energies of $3\times3\times3$ supercell with those of $4\times4\times4$ supercell at 300K and 1000K, and find that the energy fluctuations between $3\times3\times3$ and $4\times4\times4$ supercells are less than 0.01\%.  So the adequate supercell size consisting of $3\times3\times3$ unit cells is chosen to calculate thermal properties for all calculated intermetallics; however it has to be considered that Grabowski et al \cite{Grabowski} reported that small error in phonon calculation may bring significant change in the calculated thermal properties, especially at high temperatures. The thermal electronic  energies and entropies are evaluated using one-dimensional integrations from the self-consistent DFT calculations of electronic DOS using FD smearing as shown in Eqs. (\ref{ele-Energy}) and (\ref{ele-entropy}). In order to get the temperature dependence of lattice parameters, we have calculated total free energy at temperature points with a step of 1K from 0 to 1000K  at 15 volume points. At each temperature point, the equilibrium volume $V_{T}$ and isothermal bulk moduli B(T) are obtained by minimizing free-energy with respect to $V$ from fitting the integral form of the Vinet equation of state (EOS) \cite{Vinet} at $p=0$.  While,  $C_{p}$ has be calculated  by polynomial fittings for $C_{\mathrm{V}}$ and by  numerical differentiation for $\partial V / \partial T$ to obtain $\alpha (T)$.

\section{Results and discussions}
\subsection {Phonon-dispersion curves and phonon density of states}
\begin{figure}
\scalebox{0.4}[0.4]{\includegraphics{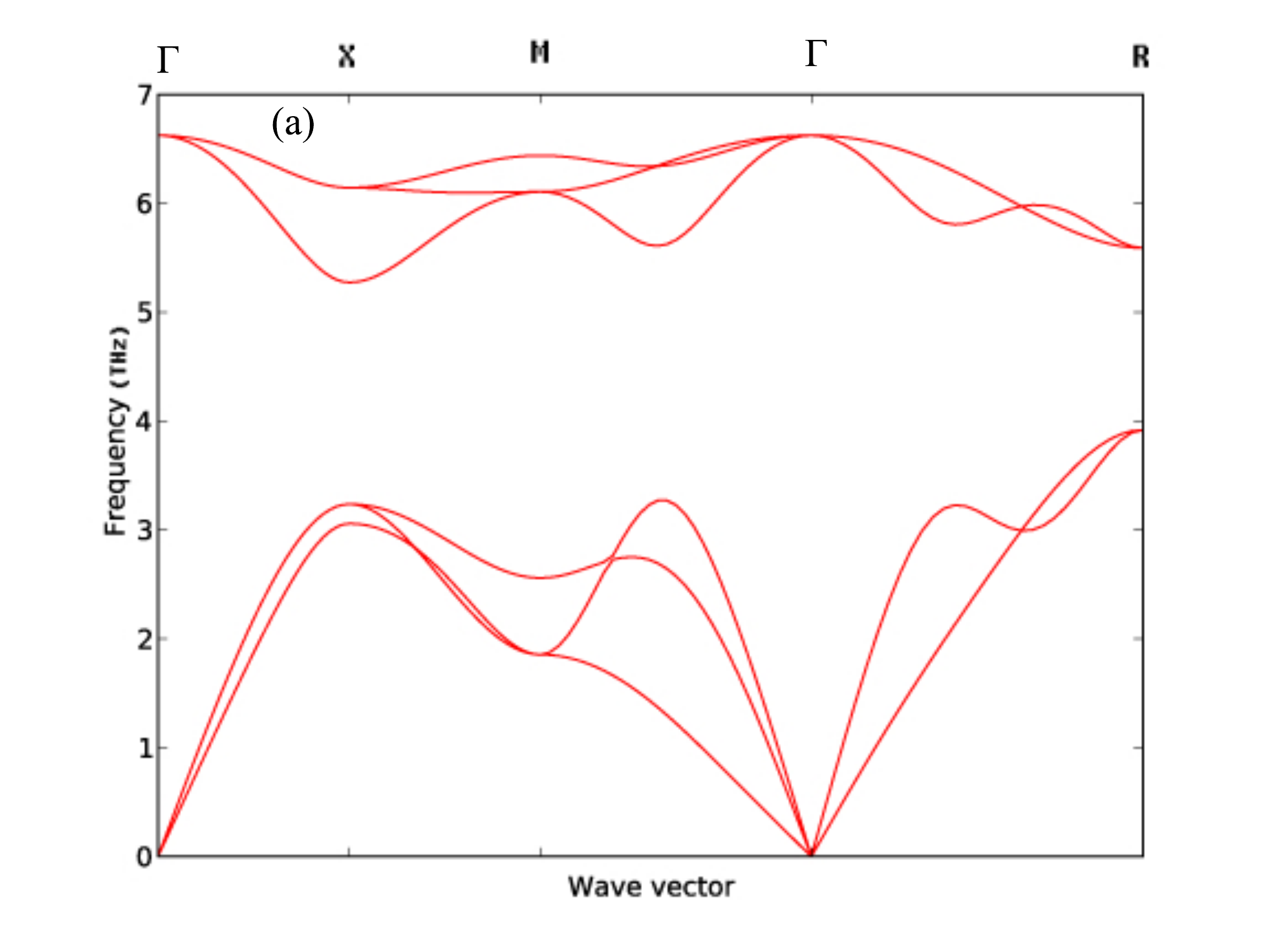}}
\scalebox{0.4}[0.4]{\includegraphics{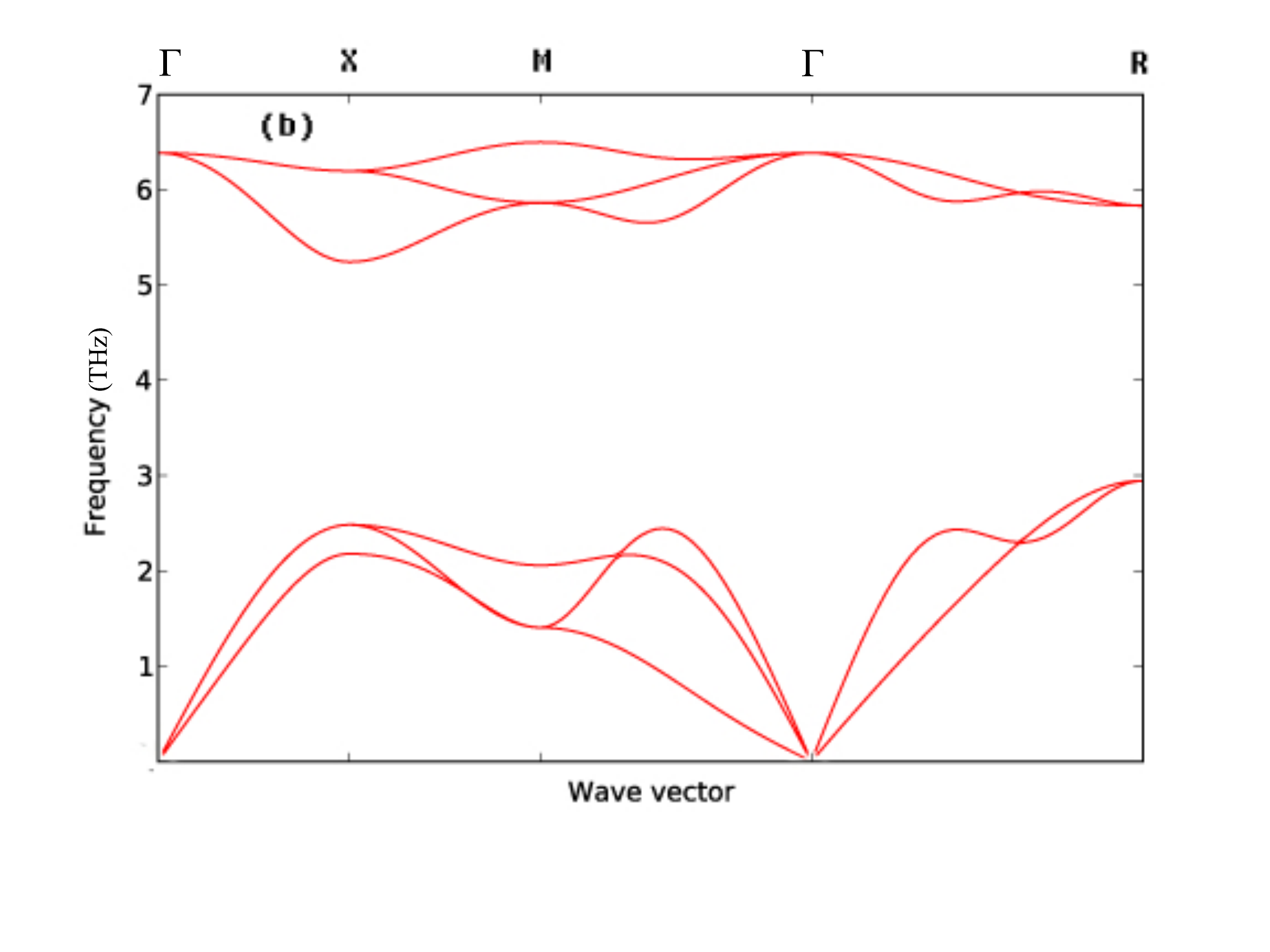}}
\scalebox{0.4}[0.4]{\includegraphics{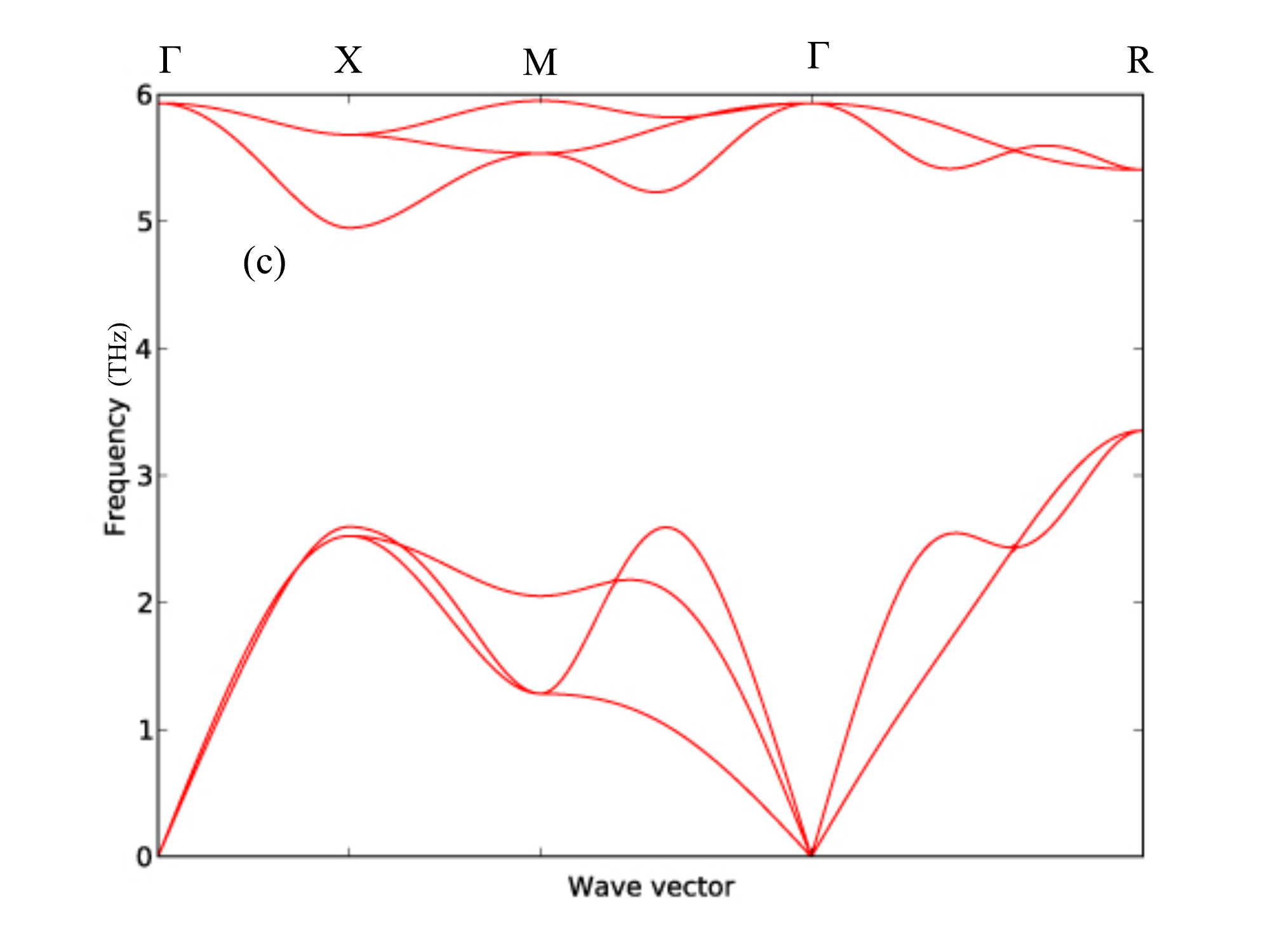}}
\scalebox{0.4}[0.4]{\includegraphics{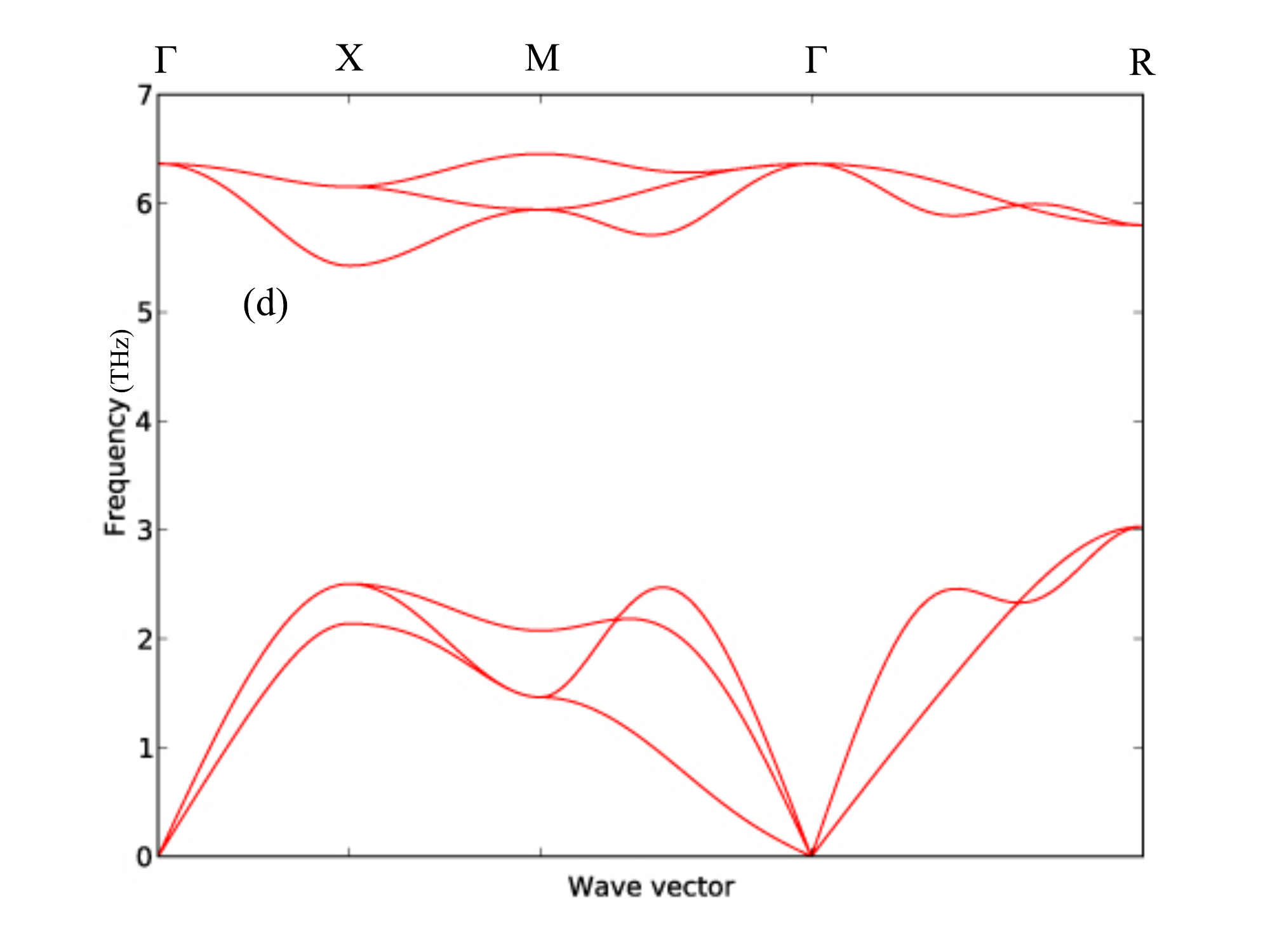}}
\caption{(Color online) Phonon-dispersion curves of (a) MgY, (b) MgDy, (c) MgPr, and (d) MgTb. }
\label{phonon}
\end{figure}

\begin{figure}
\scalebox{0.4}[0.4]{\includegraphics{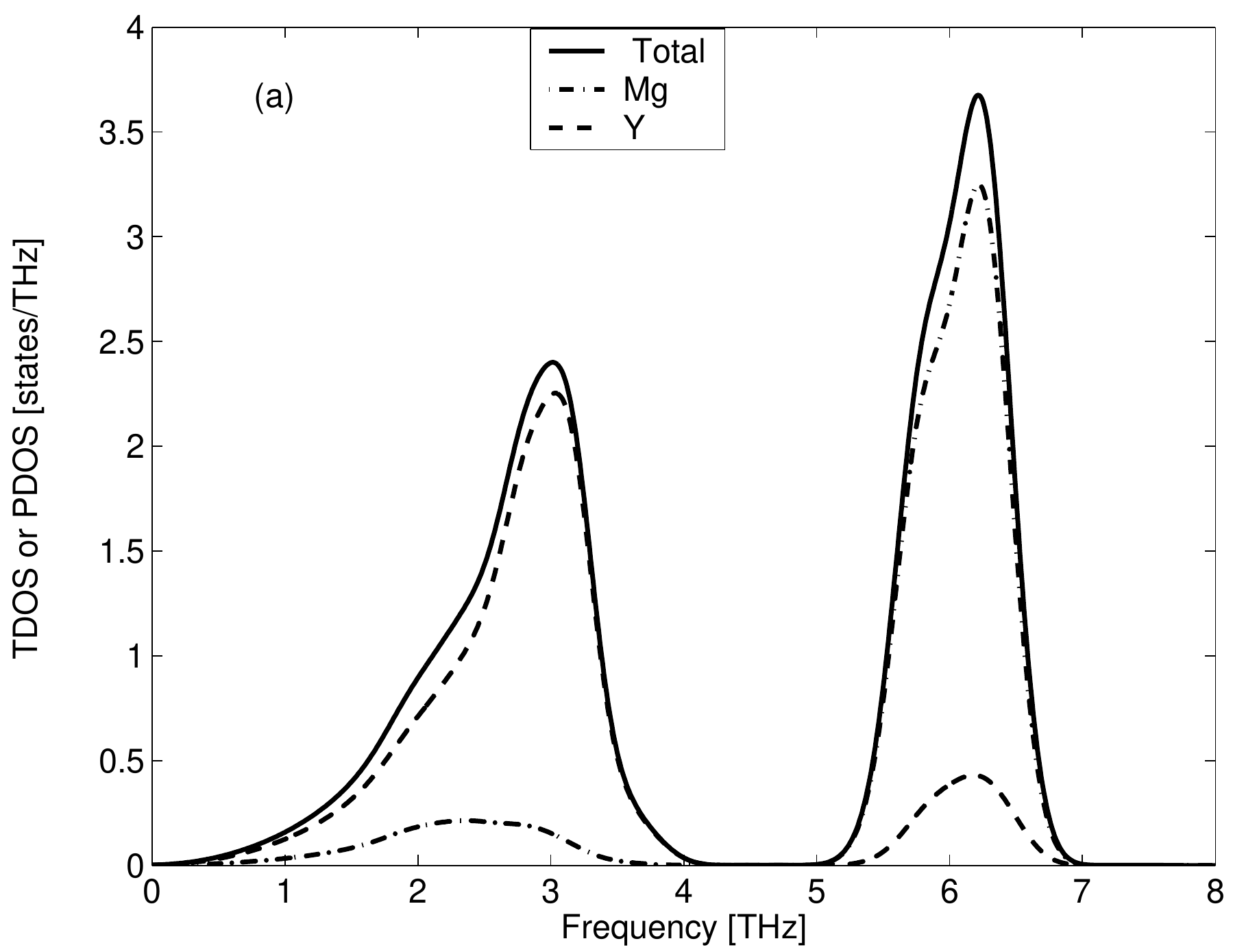}}
\scalebox{0.4}[0.4]{\includegraphics{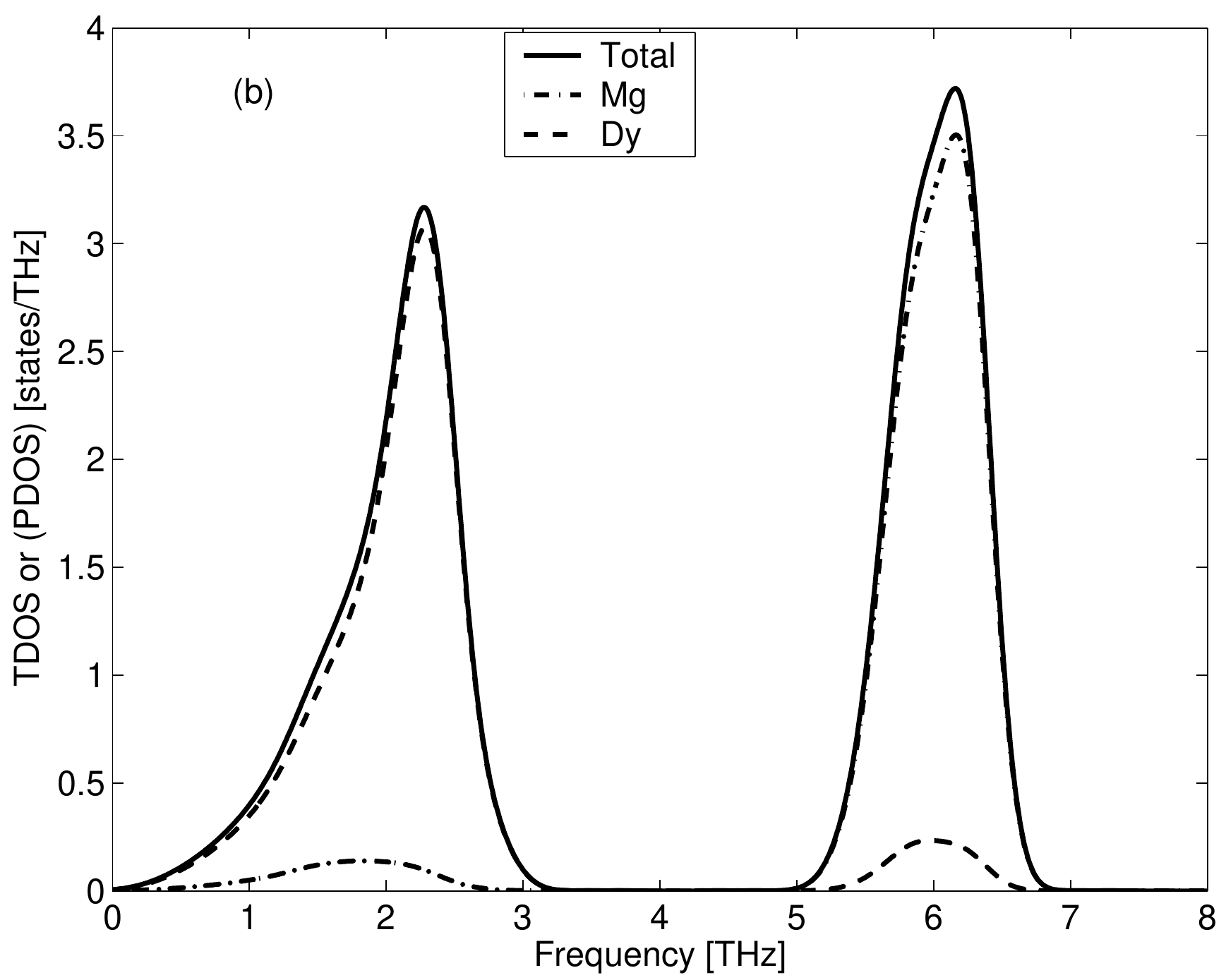}}
\scalebox{0.4}[0.4]{\includegraphics{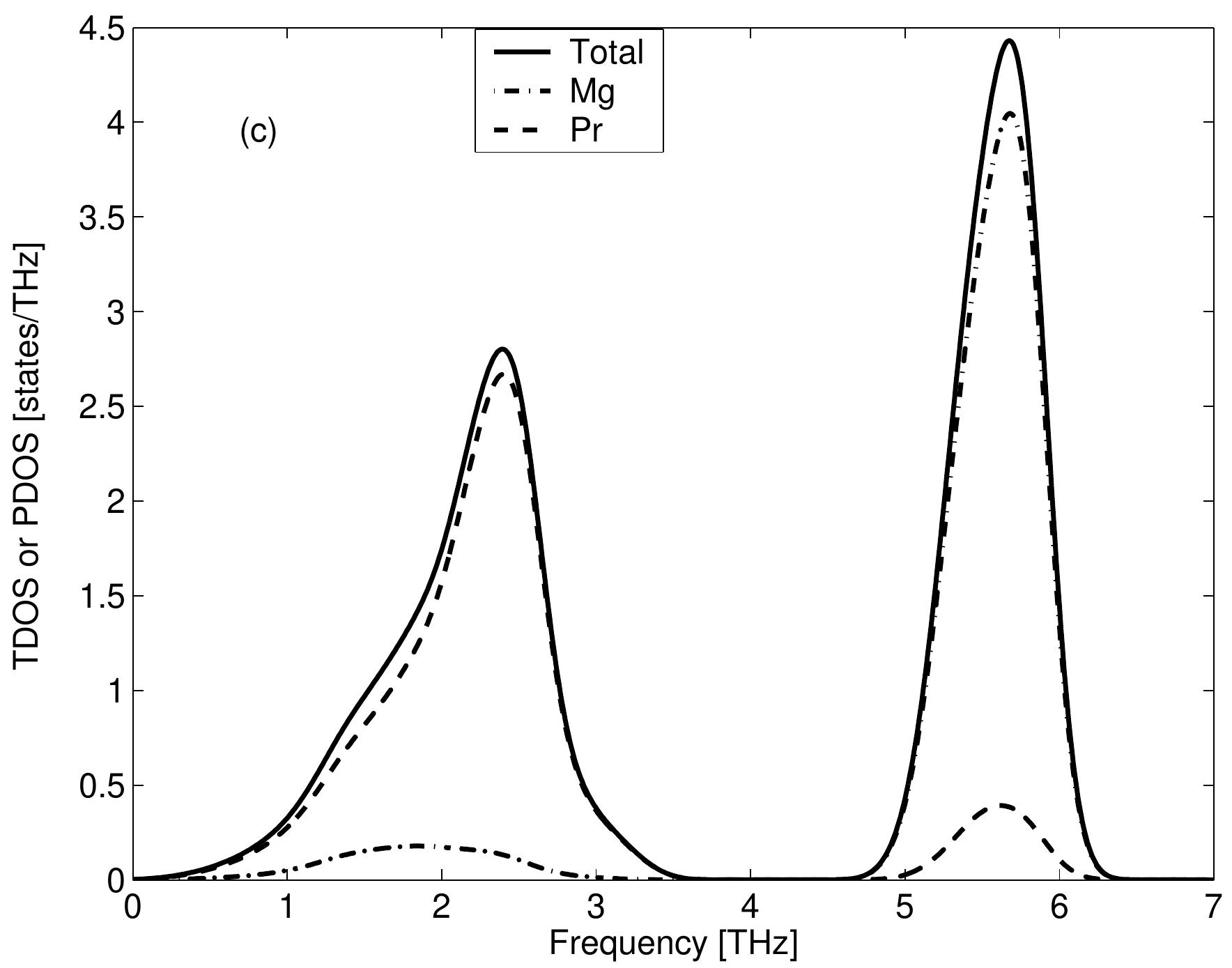}}
\scalebox{0.4}[0.4]{\includegraphics{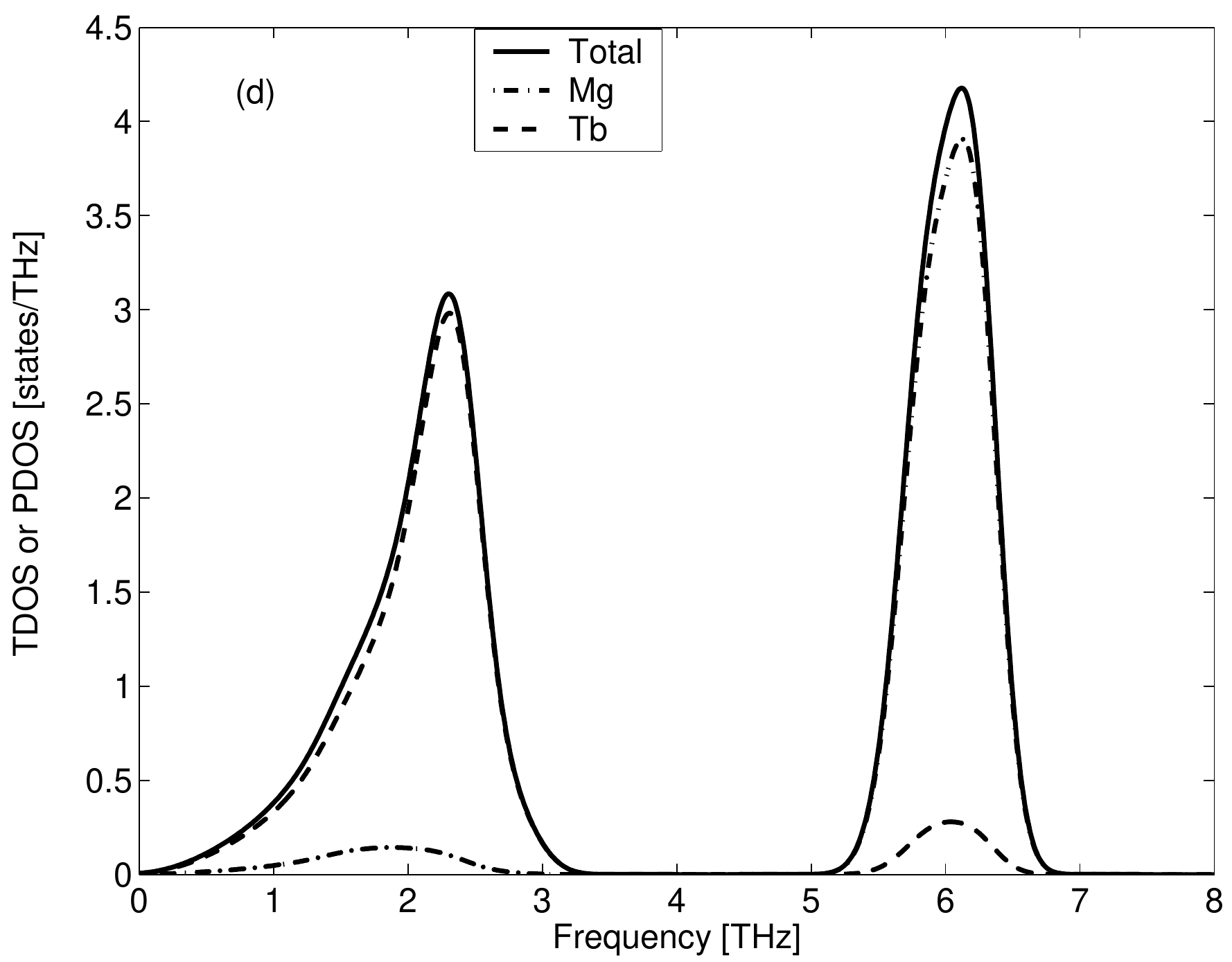}}
\caption{Phonon total density of states (TDOS) and partial density of states (PDOS) of (a) MgY, (b) MgDy, (c) MgPr, and (d) MgTb. The PDOS indicates that the density of states are mostly composed of Mg states at high frequency and RE states at low frequency.}
\label{dos}
\end{figure}

The calculated phonon-dispersion curves of four intermetallic compounds, which were calculated using $3\times3\times3$ supercells, are displayed in Figure \ref{phonon}. The phonon properties of all calculated intermetallics are computed within the generalized gradient approximation (GGA) in the B2-type structure, with space group symmetry Pm3m(221), in which the Mg atom is positioned at (0, 0, 0) and the RE (RE=Y, Dy, Pr, Tb) atoms at (0.5,0.5,0.5). These materials contain two atoms per primitive cubic unit cell. Due to the symmetry, the dispersion curves are showed along high-symmetry direction $\Gamma-X-M-\Gamma-R$ of the Brillouin zone. These dispersion curves have common framework. There is a phonon band gap starting around 3 THz. The maximum value of acoustic modes for MgY is slightly greater than 3 THz and those for MgDy, MgTb, and MgPr are slightly lighter than this value, since the atomic mass of Y is smaller than those of Dy, Tb, and Pr. The phonon density of states (DOS) including the partial DOS (PDOS) and the total DOS (TDOS) are shown in Figure \ref{dos}. Sampling a $51\times51\times51$ Monkhorst-Pack grid for phonon wave vectors $\mathbf{q}$ is found to be sufficient in order to get the mean relative error in each channel of phonon DOS. The flat regions of phonon-dispersion curves, which correspond to the peaks in the phonon PDOS, indicate localization of of the states, i.e., they behave like "atomic states" \cite{Togo}. For the four intermetallics, we find that the DOS are mostly composed of Mg states above the phonon band gap since its atomic mass is lighter than those of the rare earth elements RE (RE=Y, Dy, Pr, Tb). Our results agree with the previous theoretical investigation in which the contribution of RE atoms is dominant in phonon frequency since they are heavier than Mg atoms \cite{Wu1}.

\subsection {Bulk properties and thermal expansion}
\begin{figure}
\scalebox{0.7}[0.7]{\includegraphics{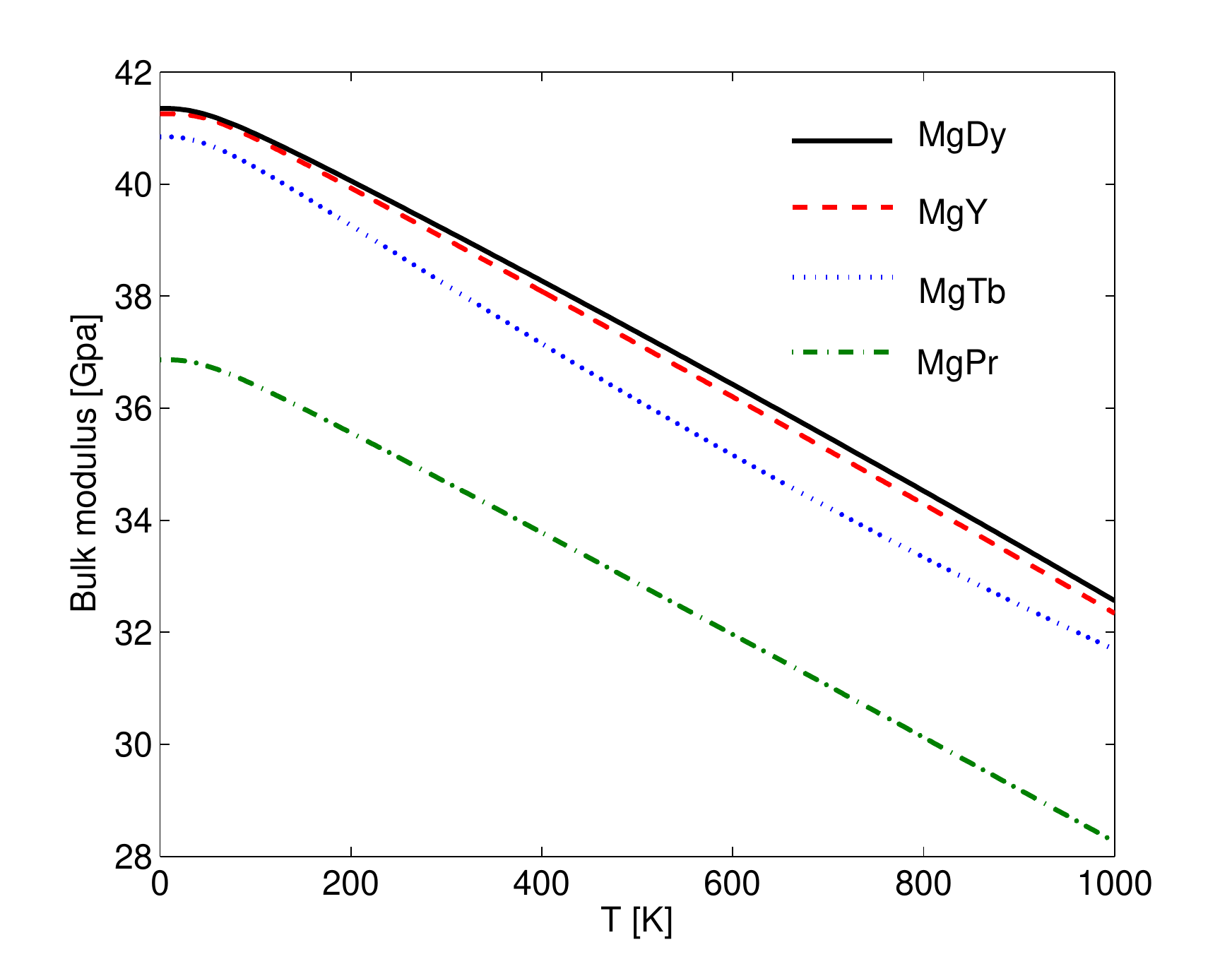}}
\caption{(Color online) Isothermal bulk moduli as a function of temperature. }
\label{Bt}
\end{figure}

The present results of the equilibrium lattice constants $a_{0}$ and isothermal bulk modulus $B_{0}$ at $T=0K$ for MgRE (RE=Y, Dy, Pr, Tb) intermetallics together with the available experimental values \cite{Villars} and the previous calculated results \cite{Wu} are shown in Table \ref{table}. Our calculated results for the equilibrium lattice constants are within 0.8\% of the experimental values, and shows excellent agreements with the previous theoretical results. For the bulk modulus $B_{0}$ which are obtained from Vinet equation of state, the present results are within 1.9\% errors from the previous calculated values. The isothermal bulk modulus as a function of temperature $B(T)$ are shown in Figure \ref{Bt}. Among the four intermetallic compounds, through the temperature range, heavy RE MgDy and light RE MgPr have the highest and lowest bulk moduli, respectively, i.e, MgDy is the most incompressible and MgPr is the most compressible. This indicates that the mechanical properties of MgRE intermetallics can be further improved by addition of heavy RE metals \cite{Wu1}.  With increasing temperature, the bulk moduli of the four intermetallics decrease and the differences among the bulk moduli almost remain unchanged. This character demonstrates that rare-earth-magnesium MgRE intermetallics can keep their mechanical properties through wide temperature range and have good high temperature stability, and it agrees well with recent reports that the MgRE have good strength at high temperature in comparison with the traditional Mg alloys \cite{Mordike1,Mordike2,Lorimer}.

\begin{figure}
\scalebox{0.7}[0.7]{\includegraphics{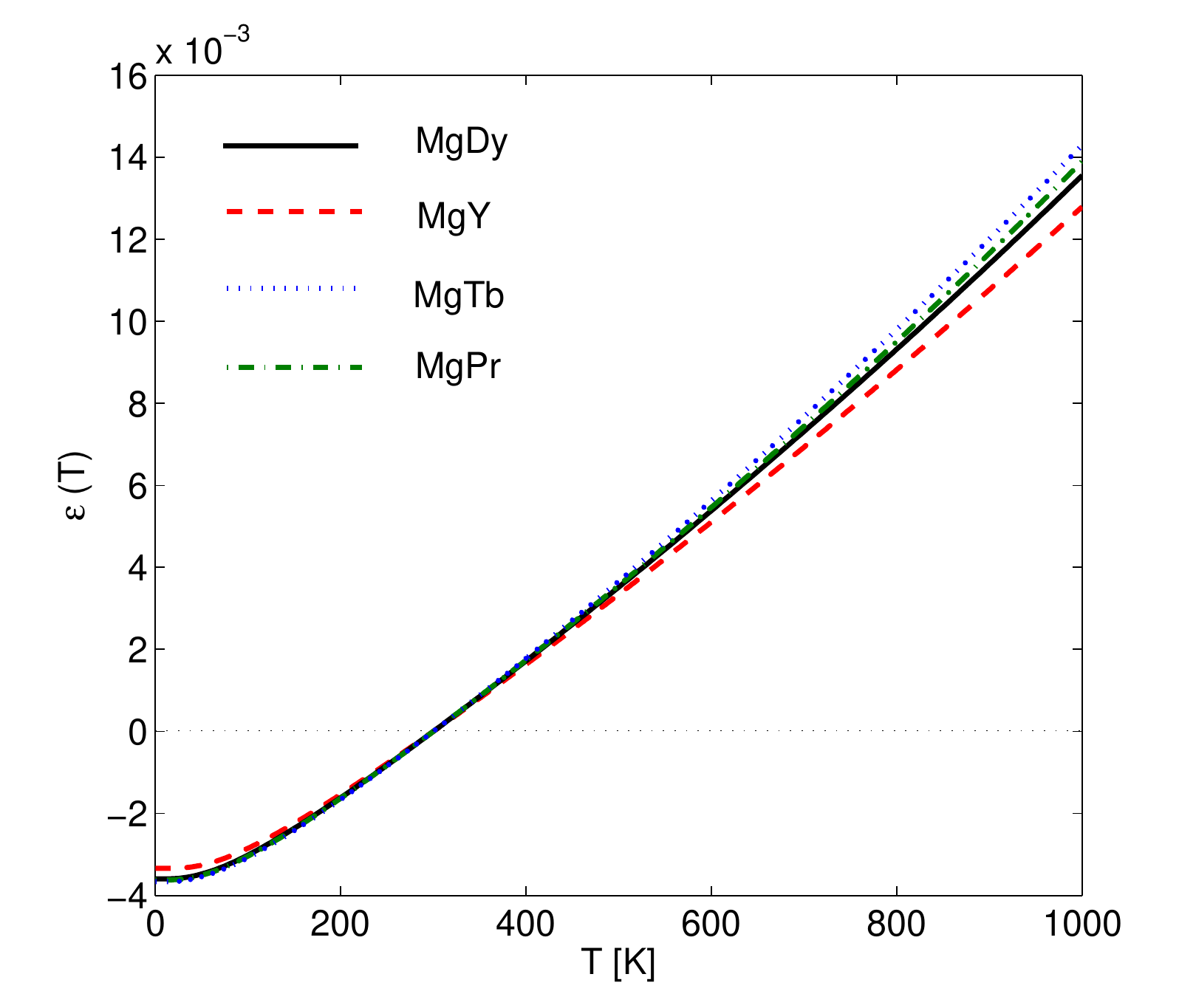}}
\caption{(Color online) Temperature dependence of the linear thermal expansion $\epsilon(T)$.}
\label{epsilonfig}
\end{figure}

\begin{figure}
\scalebox{0.7}[0.7]{\includegraphics{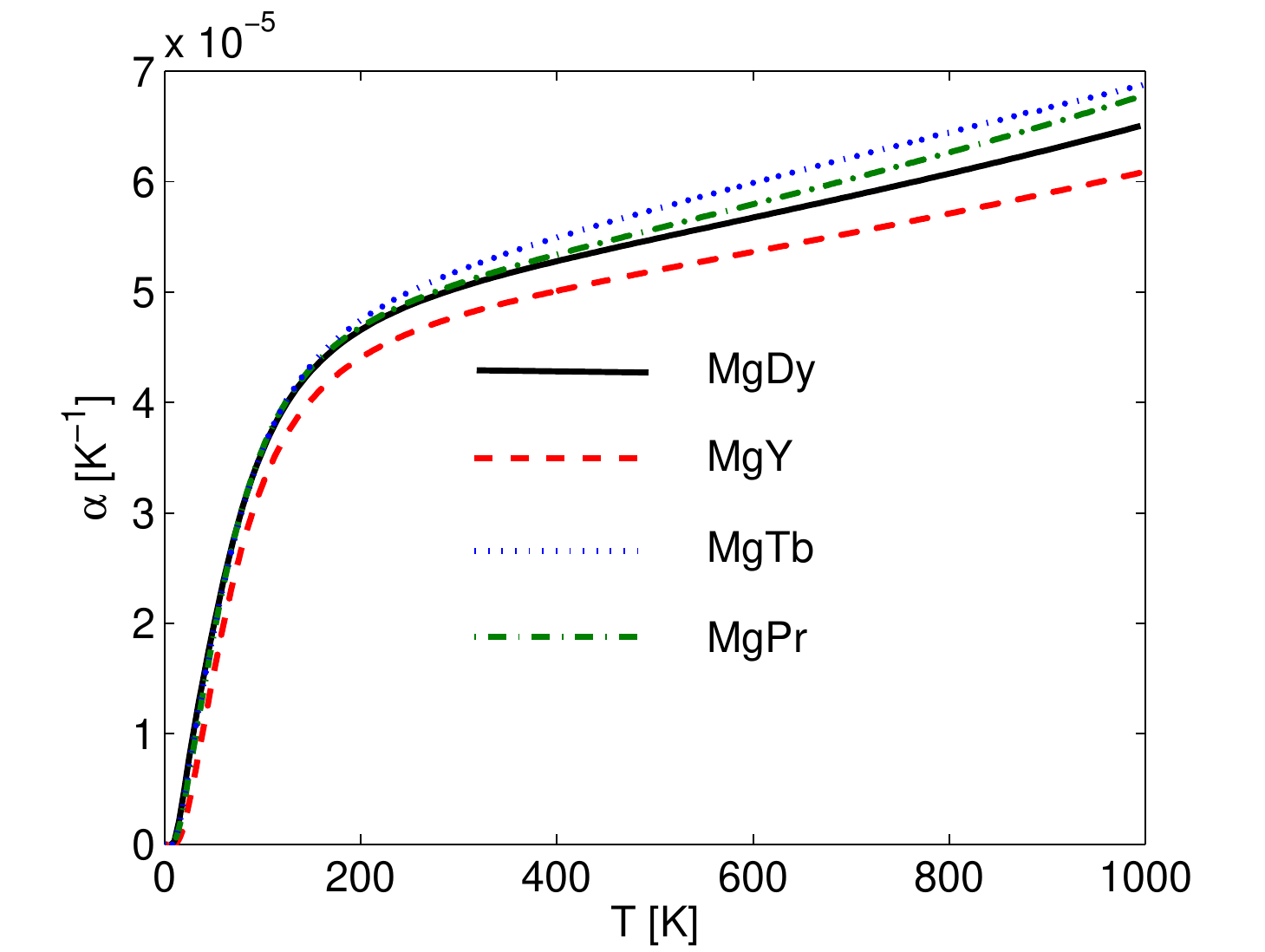}}
\caption{(Color online) The coefficients of volume thermal expansion $\alpha$ as a function of temperature.}
\label{alpha}
\end{figure}

 The temperature dependence of the linear thermal expansion $\epsilon$ defined by Eq. (\ref{epsilon}) of the four MgRE intermetallics are shown in Figure \ref{epsilonfig}. The linear expansions of the four compounds are found MgTb$>$MgPr$>$MgDy$>$MgY, however the differences among them are small. The coefficients of the volume thermal expansion $\alpha$ as a function of temperature are shown in Figure \ref{alpha}.With increasing temperature, the thermal expansion expansion grow rapidly up to $\sim400K$, and the slops become smaller and nearly constant at high temperatures. Below $\sim200K$, those of MgDy, MgTb, and MgPr are equivalent and are greater than that of MgY. The reason can be understand from the electron configurations of the corresponding elements. We have the electron structures with Mg($3s^2$), Y($4d^1 5s^2$), Dy($4f^{10}6s^2$), Pr($4f^{3} 6s^2$), and Tb($4f^{9} 6s^2$). All the elements have two $s$ electrons, and the outermostshell of rare-earth element Y includes the $d$ electrons while the other elements with $f$ electrons. So the thermal expansion of MgY is smaller than those of MgDy, MgPr, and MgTb. At high temperatures, those of MgTb and MgPr tend to be equivalent.

\subsection {Specific heat}
\begin{figure}
\scalebox{0.4}[0.4]{\includegraphics{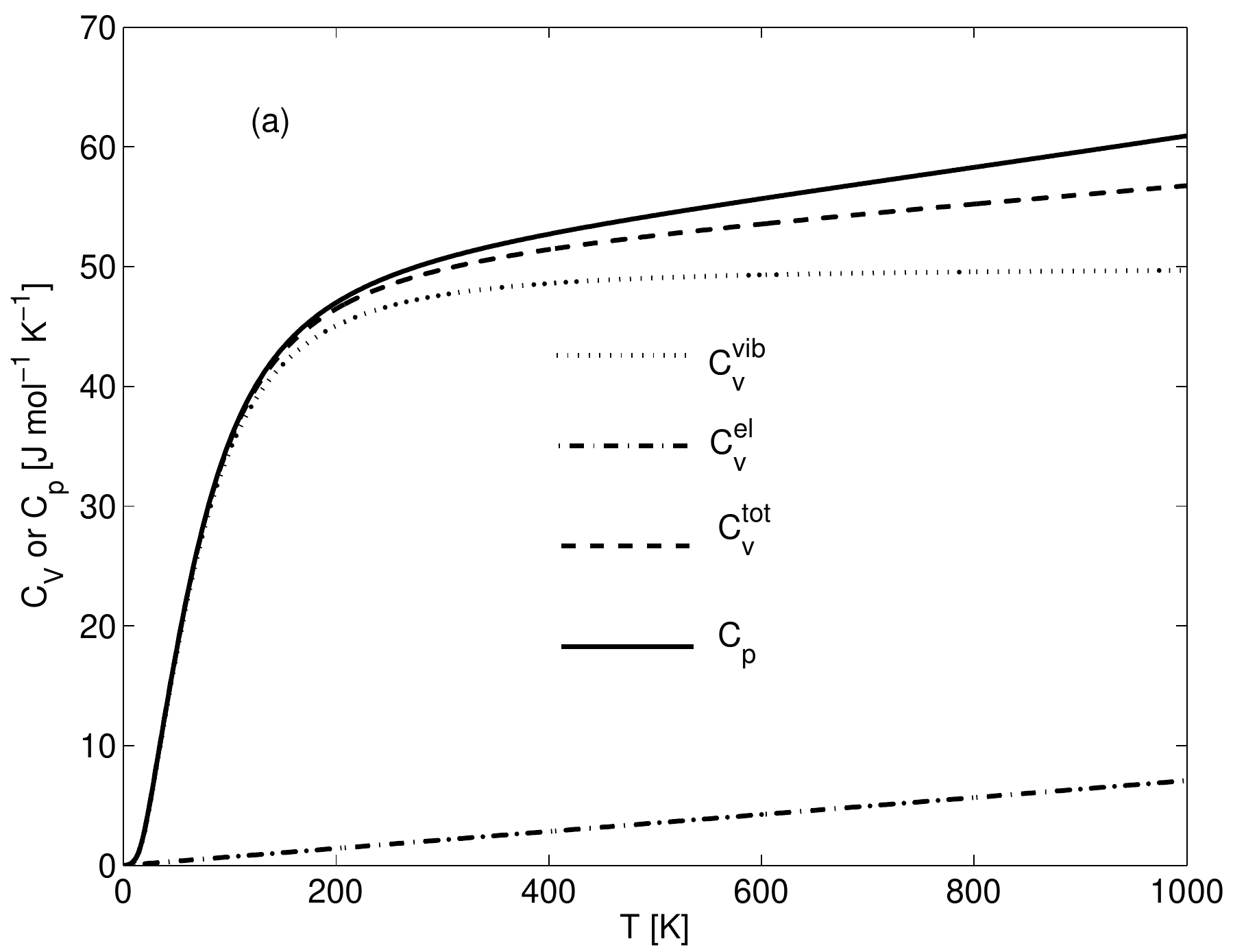}}
\scalebox{0.4}[0.4]{\includegraphics{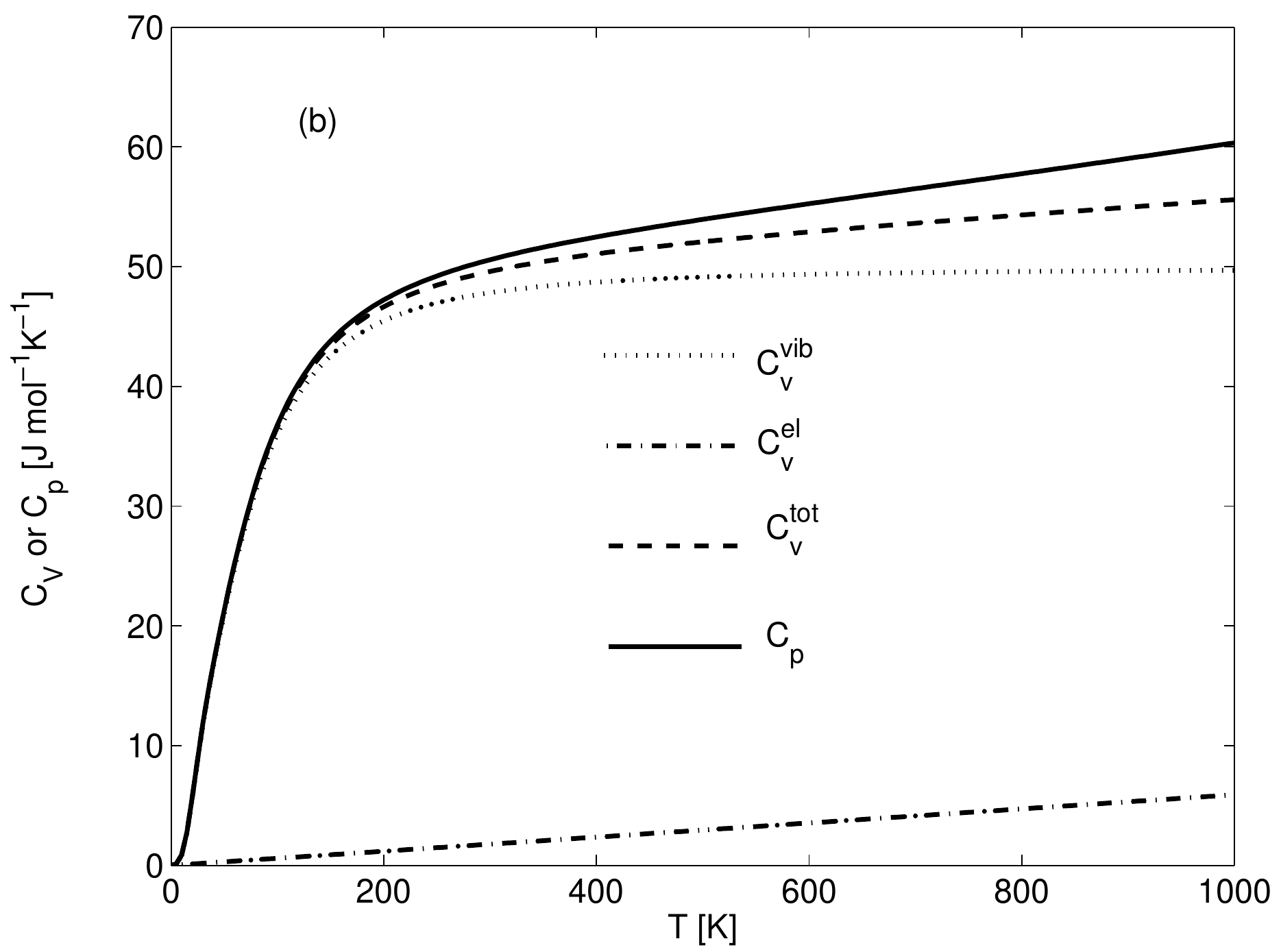}}
\scalebox{0.55}[0.55]{\includegraphics{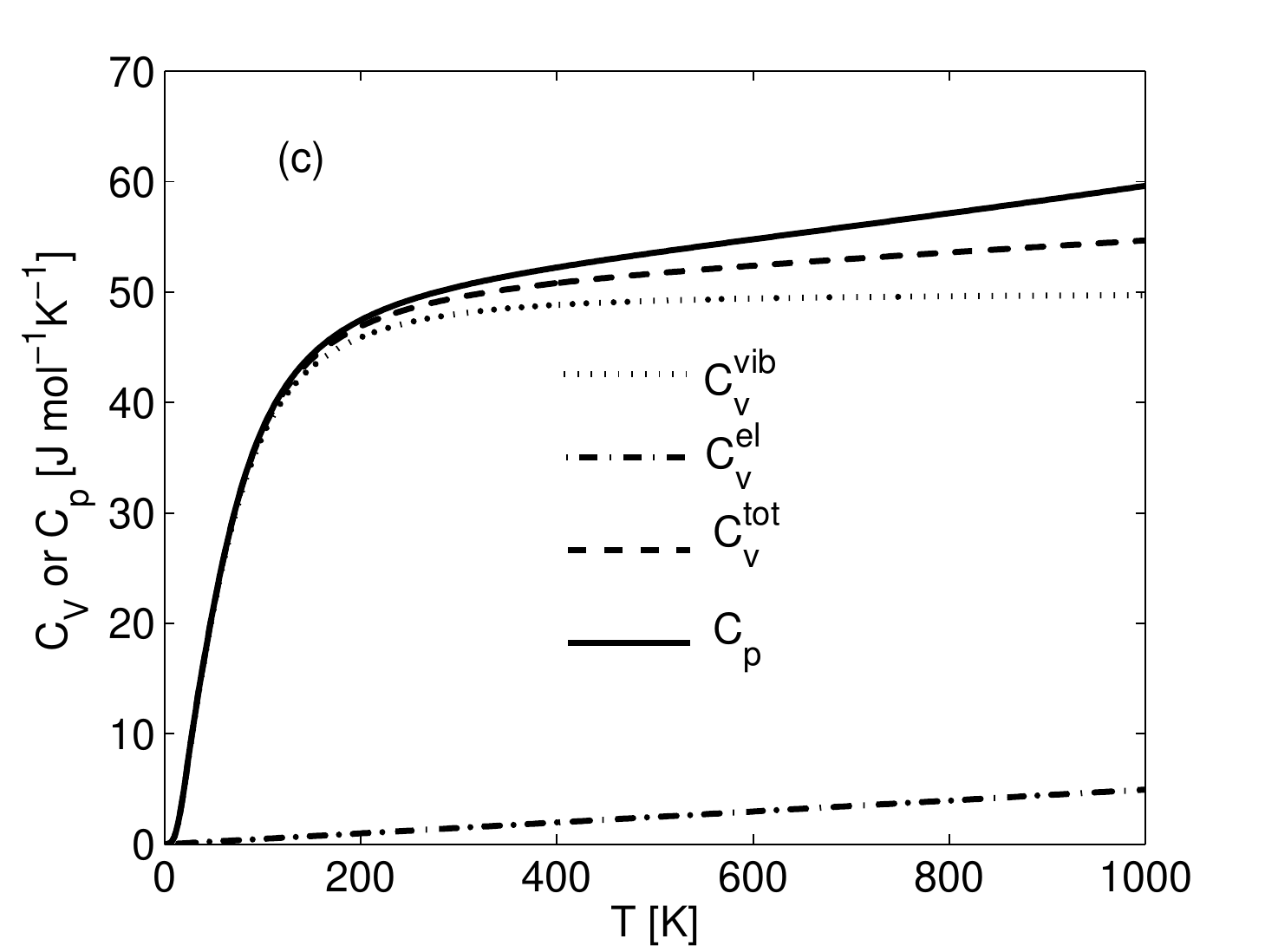}}
\scalebox{0.55}[0.55]{\includegraphics{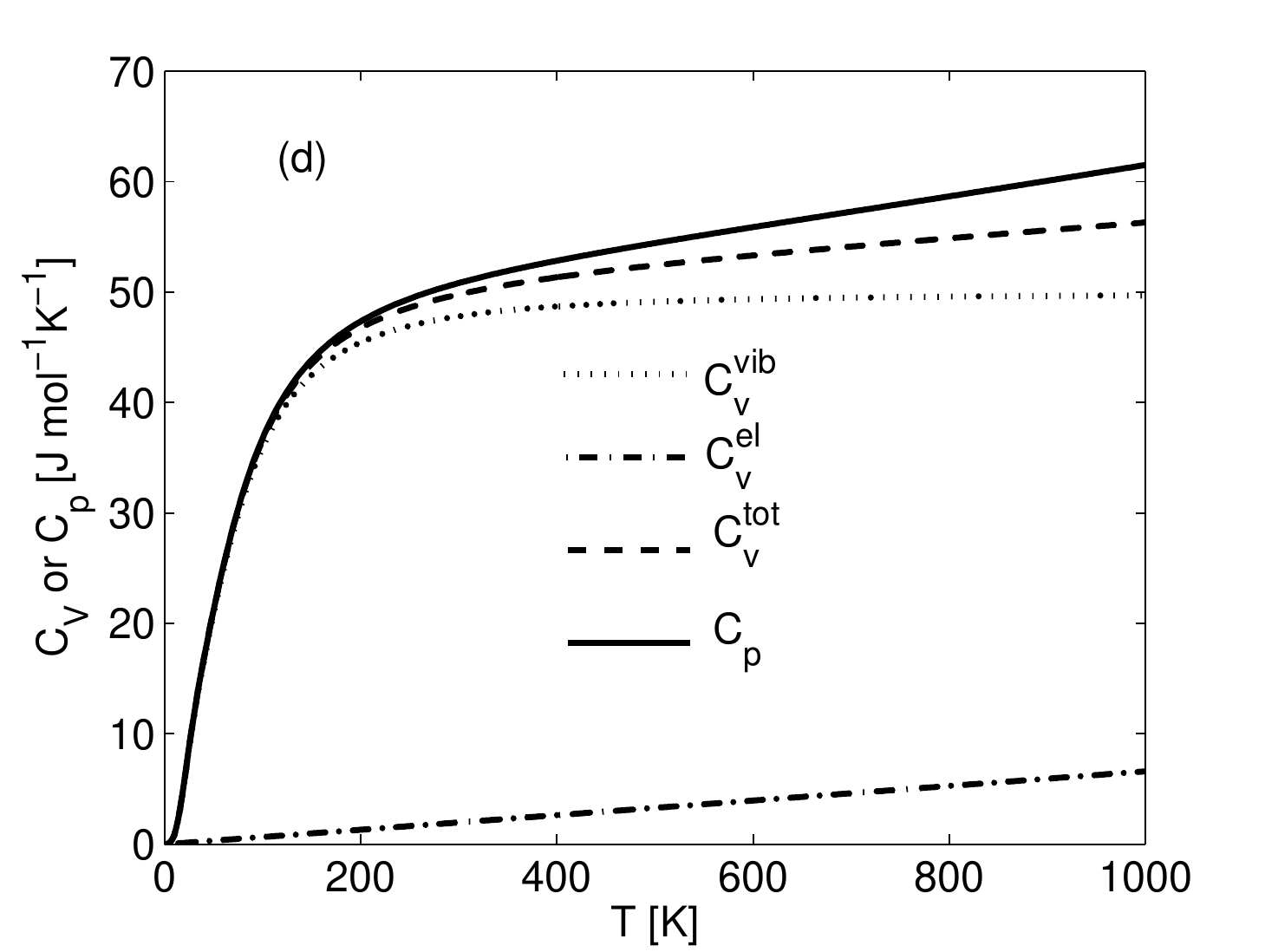}}
\caption{Temperature dependence of heat capacity of (a) MgY, (b) MgDy, (c) MgPr, and (d) MgTb. Our results shows that the thermal electronic contributions to specific heat is remarkable at high temperature.}
\label{C}
\end{figure}

Once the phonon spectrum over the entire Brillouin zone is available, the vibrational heat capacity at constant volume $C_{V}^{\mathrm{vib}}$ can be calculated by Eq. (\ref{cvib}), while the electronic contribution to heat capacity at constant volume $C_{V}^{\mathrm{el}}$ can be obtained from the electronic DOS by using Eq. (\ref{cel}). Then, the specific heat at constant pressure $C_{\mathrm{p}}$ can be computed by Eq. (\ref{cp}). As a comparison, both $C_{V}$ values, including electronic contribution and not, are plotted. We display the results in Figure \ref{C}. As temperature increases, $C_{V}^{\mathrm{vib}}$ tends to the classical constant 6R, and $C_{V}^{\mathrm{el}}$ and $C_{\mathrm{p}}$ still increase. At low temperature about below $\sim200K$, the discrepancy between $C_{\mathrm{p}}$ and $C_{\mathrm{V}}$ can be neglected. For thermal electronic contributions to specific heat, we find $C_{V}^{\mathrm{el}}$ is not negligible at high temperature though smaller than $C_{V}^{\mathrm{vib}}$. This character can be understood from the electronic DOS at the Fermi energy level $f(\varepsilon_{F})$. For MgRE, they have large electronic DOS near Fermi energy level. When we consider the electronic contribution, the theoretical $C_{\mathrm{p}}$ keeps positive slopes obviously. $C_{V}^{\mathrm{el}}$ is smaller than $C_{V}^{\mathrm{vib}}$, but it is larger than the value $C_{\mathrm{p}}-C_{\mathrm{V}}=\alpha^{2}BVT$. Usually, the electronic heat capacity can expressed as $C_{V}^{\mathrm{el}}=\gamma T$, where $\gamma$ is known as the electronic constants. To be guide to eye, we should mention that the order of these quantities from lower to higher is $\gamma(\mathrm{MgPr})<\gamma(\mathrm{MgDy})<\gamma(\mathrm{MgTb})<\gamma(\mathrm{MgY})$. As previous studied by Wu et al \cite{Wu}, the Fermi energy level for MgRE occurs above a peak in the electronic DOS; the bonding states are full, and filling of the anti-bonding states is sensitive to deviations in the local structure that affect the Fermi energy, so the electronic excitations affecting the thermal properties is remarkable. From Figure \ref{C}, it is clear that the electronic contribution should be considered for MgRE intermetallics.

\section {Conclusions}
In conclusion, thermal properties of  intermetallics MgRE (RE=Y, Dy, Pr, Tb) with B2 structures, such as thermal expansions, bulk modules, and heat capacities at constant volume and constant pressure as a function of temperature are studied by using the density functional theory and density functional perturbation theory in combination with the quasiharmonic approximation. It is found that the contribution of RE atoms is dominant in phonon frequency since Mg atomic mass is lighter than those of the rare earth elements RE.  Through the temperature range, heavy RE MgDy is the most incompressible and light RE MgPr is the most compressible. With increasing temperature, the bulk moduli of the four intermetallics decrease and the differences among the bulk moduli almost remain unchanged. In the thermal expansion, we find MgTb$>$MgPr$>$MgDy$>$MgY at hight temperature, and below $\sim200K$ those of MgDy, MgTb, and MgPr are equivalent and are greater than that of MgY.  The electronic electronic contributions to the specific heat are discussed, and found to be important for the calculated MgRE intermetallics.


 \vskip 2in

\footnotesize

\def\refname{{\large\bfseries References}}

\newpage
\begin{table}
\caption{The equilibrium lattice constants $a_{0}$ and bulk modulus $B_{0}$ at $T=0K$ for MgRE (RE=Y, Dy, Pr, Tb) in our calculation in comparison with the previous calculated results and the experiment.}
\begin{tabular}{ccccc}
  \hline
   & MgDy& MgY & MgTb & MgPr \\
   \hline
  $a_{0}$ & 3.778, 3.765$^{a}$, 3.759$^{b}$ & 3.795, 3.796$^{a}$, 3.796$^{b}$& 3.789, 3.781$^{a}$, 3.781$^{b}$ & 3.910, 3.901$^{a}$, 3.912$^{b}$ \\
  $B_{0}$ & 41.35, 42.32$^{a}$ & 41.25, 42.06$^{a}$ & 40.84, 41.89$^{a}$ & 36.86, 37.20$^{a}$ \\
  \hline
\end{tabular}

\begin{tabular}{cccccccc}
  \leftline {${}^{a}$Reference\cite{Wu};}\\
  \leftline {${}^{b}$Reference\cite{Villars},\ \ experiment.}\\
\end{tabular}
\label{table}
\end{table}

\begin{thebibliography}{5}
\bibitem{Kainer} K. U. Kainer, Magnessium Alloys and Their
Applications(Weinheim: Wiley-VCH)(2000).
\bibitem{Mordike1} B. L. Mordike, Mater. Sci. Eng. A 324( 2002) 103.
\bibitem{Mordike2} B. L. Mordike, J. Mater. Process. Technol. 117 (2001)
381.
\bibitem{Lorimer} G. W. Lorimer, P. J. Apps, H. Karimzaden, J. F. King,
Mater. Sci. Forum 419-422 (2003) 279.
\bibitem{Luca} S. E. Luca, M. Amara, R. M. Galera, J. F. Berar, J.
 Phys:Condens. Matter 14 (2002) 935.
\bibitem{Deldem} M.Deldem, M. Amara, R. M. Galera, P. Morin, D. Schmitt,
B. Ouladdiaf, J. Phys:Condens. Matter 10 (1998) 165.
\bibitem{Wu} Y. Wu, W. Hu, Eur.Phys. J. B 60 (2007) 75.
\bibitem{Wang} R. Wang, S. F. Wang, X. Z. Wu, Y. Yao, and A. P. Liu, Intermetallics 18 (2010) 2472.
\bibitem{Wang1} X. Z. Wu, R. Wang, S. F. Wang, and L. L. Liu, Physica B 406 (2011) 967.
\bibitem{Tao} X. M. Tao, Y. F. Ouyang, H. S. Liu, Y. P. Feng, Y. Du, Z.P. Jin, Solid State Communications 148 (2008) 314.
\bibitem{Guo} C. Guo, Z. Du, J. Alloys Compounds 422 (2006), p. 102.
\bibitem{Cacciamani} G. Cacciamani, S. De Negri, A. Saccone, R. Ferro, Intermetallics 11 (2003)
1135.
\bibitem{Wu1} Y. Wu, W. Hu, J. Phys. D: Appl. Phys. 40 (2007) 7584.
\bibitem{Moruzzi} V. L. Moruzzi, J. F. Janak, and K. Schwarz, Phys. Rev. B 37 (1988) 790.
\bibitem{Biernacki} S. Biernacki and M. Scheffler, Phys. Rev. Lett. 63 (1989) 290 .
\bibitem{Pavone} P. Pavone, K. Karch, O. Sch¨¹tt, W. Windl, D. Strauch, P. Giannozzi,
and S. Baroni, Phys. Rev. B 48 (1993) 3156 .
\bibitem{Carrier} P. Carrier, R. Wentzcovitch, and J. Tsuchiya, Phys. Rev. B 76  (2007)
064116.
\bibitem{Nie} Y. Nie, Y. Xie, Phys Rev B 75 (2007) 174117.
\bibitem{Togo} A. Togo, L. Chaput, I. Tanaka, G. Hug, Phys Rev B 81 (2010) 174301.
\bibitem{Giannozzi} P. Giannozzi, S. de Gironcoli, P. Pavone, S. Baroni S. Phys Rev B 43 (1991) 7231.
\bibitem{Wang2004} Y. Wang, Z. K. Liu, L. Q. Chen, Acta. Mater. 52 (2004) 2665.
\bibitem{Karki} B. B. Karki, R. M. Wentzcovitch, S. de Gironcoli, and S. Baroni, Phys. Rev. B 62 (2000) 14750.
\bibitem{Gurel} T. G\"{u}rel and R. Eryi\v{g}it, Phys. Rev. B 82 (2010) 104302.
\bibitem{Kresse1} G. Kresse, J. Hafner, Phys. Rev. B 48 (1993) 3115.
\bibitem{Kresse2} G. Kresse, J. Furthm¨¹ller,  Comput. Mater. Sci. 6 (1996) 15.
\bibitem{Kresse3} G. Kresse, J. Furthm¨¹ller,  Phys. Rev. B 54 (1996) 11169.
\bibitem{Blochl} P. E. Bl$\ddot{o}$chl, Phys. Rev. B 50 (1994) 17953.
\bibitem{Kresse4} G. Kresse, D. Joubert, Phys. Rev. B 59 (1999) 1758.
\bibitem{Perdew1} J. P. Perdew, K. Burke, M. Ernzerhof, Phys. Rev. Lett. 77 (1996) 3865.
\bibitem{Perdew2} J. P. Perdew, K. Burke, M. Ernzerhof, Phys. Rev. Lett. 78 (1996) 1396.
\bibitem{Monkhorst} H. J. Monkhorst, J. D. Pack, Phys. Rev. B 13 (1976) 5188.
\bibitem{Methfessel} M. Methfessel, A. T. Paxton, Phys. Rev. B 40 (1989)3616.
\bibitem{Kresse5} G. Kresse, M. Marsman, J. Furthm\"{u}ller, VASP the guide, http://cms.mpi.univie.ac.at/vasp/.
\bibitem{Togo2008}A. Togo, F. Oba, I. Tanaka, Phys. Rev. B, 78 (2008) 134106.
\bibitem{Togop} A. Togo, Phonopy, http://phonopy.sourceforge.net/.
\bibitem{Grabowski} B. Grabowski, L. Ismer, T. Hickel, J. Neugebauer, Phys. Rev. B 79 (2009) 134106.
\bibitem{Vinet} P. Vinet, J. H. Rose, J. Ferrante, J. R. Smith, J. Phys.: Condens. Matter 1 (1989) 1941.
\bibitem{Villars} P. Villars, L. D. Calvert, Pearson's Handbook of
Crystallographic Data for Intermetallic Phases (ASM, Metals Park,
oH, 1985).


\end{thebibliography}
\end{document}